\documentclass[sigconf]{acmart}

\usepackage{booktabs} 

\copyrightyear{2019} 
\acmYear{2019} 
\setcopyright{acmlicensed}
\acmConference[HotMobile '19]{The 20th International Workshop on Mobile Computing Systems and Applications}{February 27--28, 2019}{Santa Cruz, CA, USA}
\acmBooktitle{The 20th International Workshop on Mobile Computing Systems and Applications (HotMobile '19), February 27--28, 2019, Santa Cruz, CA, USA}
\acmPrice{15.00}
\acmDOI{10.1145/3301293.3302377}
\acmISBN{978-1-4503-6273-3/19/02}

\hypersetup{draft}

\begin{document}
\title[Earthquake Early Warning and Beyond: Systems Challenges in ...]{Earthquake Early Warning and Beyond: Systems Challenges in Smartphone-based Seismic Network}

\author{Qingkai Kong}
\orcid{0000-0002-7399-0661}
\affiliation{%
  \institution{University of California, Berkeley}
  \streetaddress{209 McCone Hall}
  \city{Berkeley}
  \state{CA}
  \postcode{94720-4767}
}
\email{kongqk@berkeley.edu}

\author{Qin Lv}
\affiliation{%
  \institution{University of Colorado Boulder}
  \streetaddress{430 UCB}
  \city{Boulder}
  \state{CO}
  \postcode{80309-0430}
}
\email{qin.lv@colorado.edu}

\author{Richard M. Allen}
\affiliation{%
  \institution{University of California, Berkeley}
  \streetaddress{279 McCone Hall}
  \city{Berkeley}
  \state{CA}
  \postcode{94720-4767}
}
\email{rallen@berkeley.edu}

\begin{abstract}
Earthquake Early Warning (EEW) systems can effectively reduce fatalities, injuries, and damages caused by earthquakes. Current EEW systems are mostly based on traditional seismic and geodetic networks, and exist only in a few countries due to the high cost of installing and maintaining such systems. The MyShake system takes a different approach and turns people's smartphones into portable seismic sensors to detect earthquake-like motions. However, to issue EEW messages with high accuracy and low latency in the real world, we need to address a number of challenges related to mobile computing. In this paper, we first summarize our experience building and deploying the MyShake system, then focus on two key challenges for smartphone-based EEW (sensing heterogeneity and user/system dynamics) and some preliminary exploration. We also discuss other challenges and new research directions associated with smartphone-based seismic network. 
\end{abstract}

%
%

\ccsdesc[500]{Human-centered computing~Ubiquitous and mobile computing systems and tools}
\ccsdesc[500]{Applied computing~Earth and atmospheric sciences}

\keywords{Earthquake Early Warning, Smartphone Seismic Network}

\maketitle

\section{Introduction}

Earthquakes are global hazards that frequently shake our nerves at various places on the Earth by killing people, interrupting normal life and work, and destroying cities. In order to record and understand earthquakes, instruments such as seismometers are installed globally to convert earthquake waves into digital time series including acceleration, velocity or displacement of the ground motion. Although many scientists and engineers have devoted their lives to study earthquakes, it is still not feasible to predict earthquakes using today's science and technology. The recent development of Earthquake Early Warning (EEW) systems provides at least one way to identify the occurrence of an earthquake in near real-time and issue a warning to the public \cite{Allen2009}. The effectiveness of EEW has been proved in various regions over the past decade by reducing fatalities, injuries, and damage caused by earthquakes, by alerting people to take cover, slowing down and stopping trains, opening elevator doors, and many other applications \cite{Strauss2016}. The concept of EEW is simple -- seismic waves generated by earthquakes travel at the speed of sound, while electronic signals travel at the speed of light (analogous to seeing lightning before hearing the sound of thunder). If we can detect seismic waves quickly after the earthquake occur, we can leverage electronic signals travels much faster than the seismic waves to warn people at further distances before seismic waves arrive \cite{Hiroo2005}. 

Traditional seismometers are high-quality research-grade sensors, which are costly to deploy and maintain. As such, only a limited number of seismic networks exist in the world to monitor earthquakes, and few places (will) have EEW systems (e.g., Japan, California, Taiwan, China, Mexico, Italy, Turkey, Romania, Switzerland). Many other regions with high earthquake hazards and dense populations (e.g., Nepal, Ecuador, New Zealand, Indonesia) do not have EEW systems \cite{Allen2009}. Even for places with EEW systems, many of them are limited by low station density due to the lack of funding to increase the number of sensors.  

To overcome the limitations of traditional seismic networks, the MyShake system takes a different approach -- a smartphone-based seismic network that turns people's smartphones into portable seismic sensors \cite{Kong2016_SciAdvances}. Using sensors and communication units that are readily-available in consumer smartphone devices, we can achieve rapid detection of earthquakes and issue warnings to individual users in target regions. The advantages of building such a smartphone-based EEW system are multifold: (a) no need to deploy sensors and maintain them, (b) easily scale up to the global level, (c) increase public awareness and knowledge of earthquakes. This approach also allows us to bring EEW to any region where the local population is exposed to earthquake hazards, especially in areas where do not exist the traditional EEW system. 

Such a high-gain system does come with a number of unique challenges in the real world. In this paper, we first summarize our experience building and deploying the MyShake system, then present the unique challenges of EEW and our initial exploration to address these challenges. While initiated as a seismology project, through this workshop paper, our goal is to introduce MyShake to the mobile computing community, so that we could seek expert feedback on possible solutions, potential improvements, and even new challenges/directions. 

\section{MyShake System}
MyShake is a free Android app that has the ability to recognize earthquake shaking using the sensors in every smartphone. The app runs ``silently'' in the background, and when the shaking fits the vibrational profile of an earthquake, the app sends the anonymous information to a central server, which then confirms the location and magnitude of the quake by aggregating phones in a region. The whole system design is a collaboration between academia and industry, where seismologists at Berkeley Seismology Lab provided earthquake knowledge and designed the detection algorithms, while developers from Deutsche Telecom Silicone Valley Innovation Center implemented the whole system. An upgraded version of MyShake with new UIs and functionalities for both Android and iOS phones will be released in Spring 2019 to better engage participants and start issuing earthquake early warning to the public \cite{Rochford2018_Frontiers}. In this section, we give an overview of MyShake's current status and the overall system design. 

\begin{figure}
\includegraphics[width=0.47\textwidth]{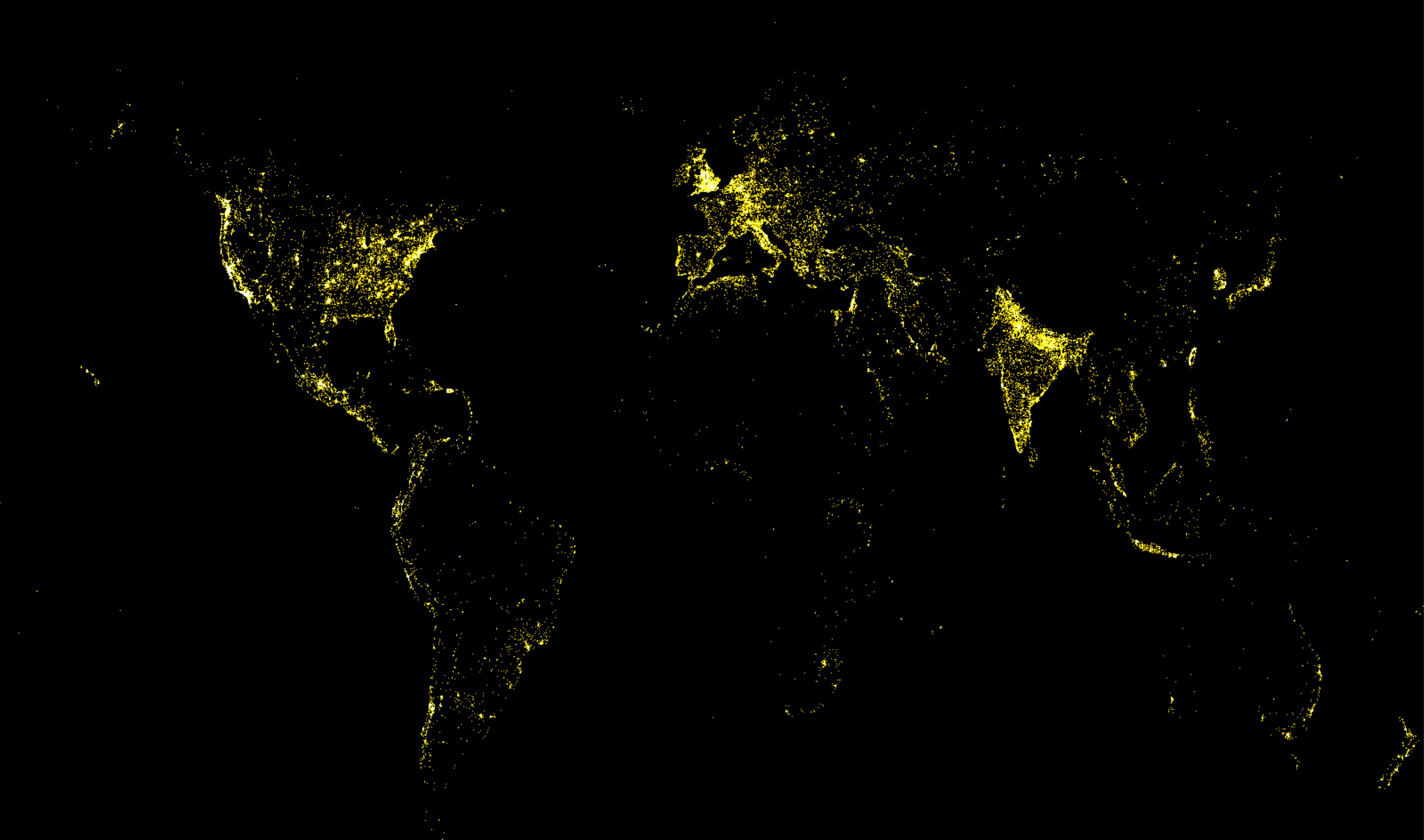}
\caption{MyShake global user distribution. Brighter color indicates higher user density. Data used here are from all registered users with locations available during the period of 2016-02-12 to 2018-08-12.}
\label{fig:myshake_user}
\vspace{-1.0em} 
\end{figure}

\subsection{MyShake Current Status}

MyShake was released to the public on 2016-02-12 and grew rapidly into a global seismic network. It currently has more than 296K downloads, 40K active users, with 6K to 7K phones contributing data on a daily basis. Figure~\ref{fig:myshake_user} shows the global distribution of MyShake users with available location information. We can see that the MyShake seismic network has already reached global coverage, and new users can join the network easily by downloading the MyShake app. Initial observations from the MyShake users show very promising results, indicating that the data collected from the phones are capable of supporting various seismological applications \cite{Kong2016_GRL}. Within the first two and half years, the MyShake network has detected around 900 earthquakes globally with magnitudes ranging from M1.6 to M7.8. There are also initial results showing that the MyShake network could potentially provide structural health monitoring of buildings \cite{Kong2018_SRL_Building}, or use the sensor array to detect smaller earthquakes.

\begin{figure}
\includegraphics[width=0.48\textwidth]{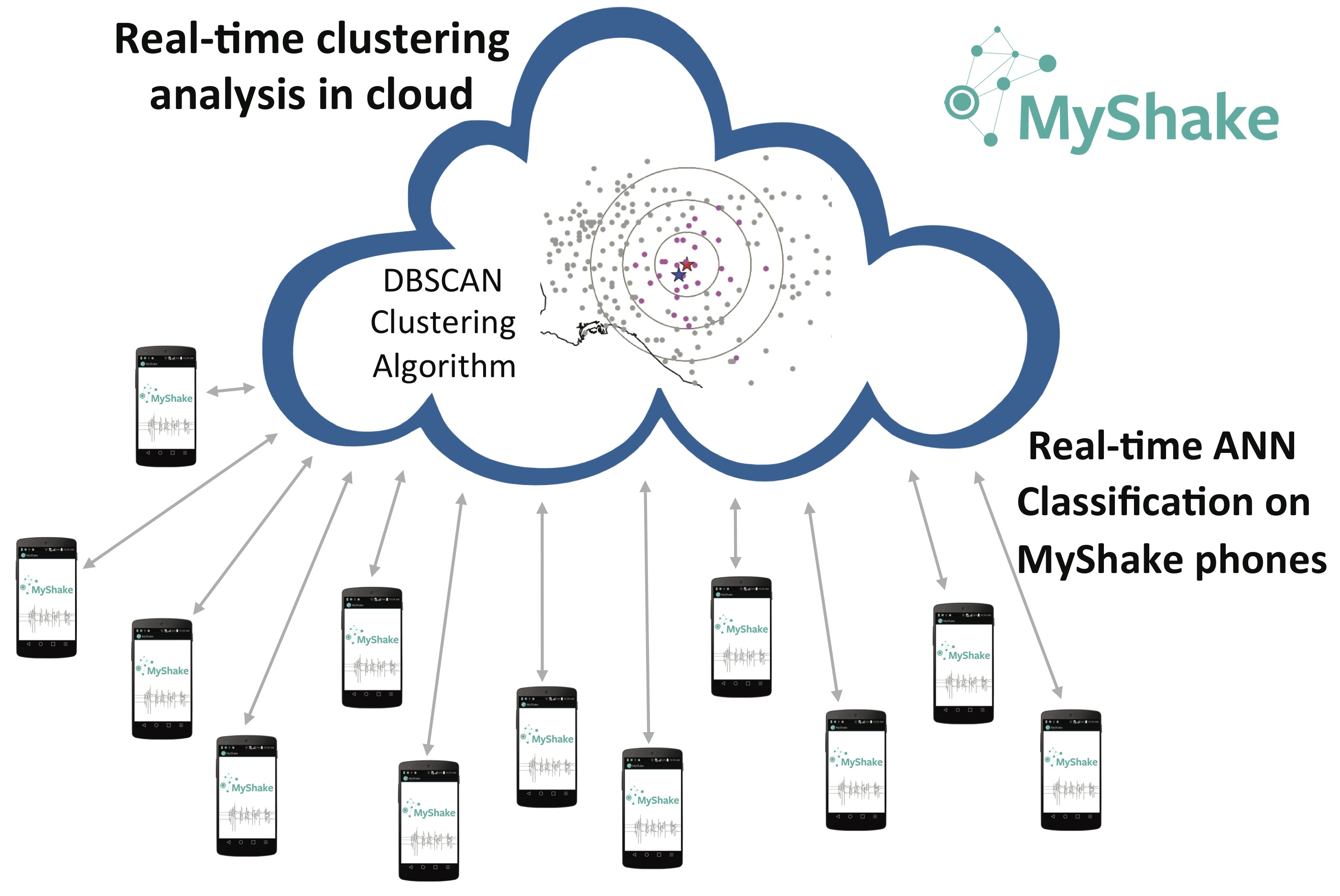}
\caption{An overview of the MyShake system.}
\label{fig:myshake_system}
\end{figure}

\subsection{MyShake System Design}

 Figure~\ref{fig:myshake_system} illustrates the current design of the MyShake system, which consists of two key components: (1) Each phone  downloads MyShake application which has the capability of listening to the accelerometer and making decisions whether the experienced motion is due to earthquake by using an artificial neural network (ANN) model. The ANN model is trained by searching for the different characteristics between earthquake and human motions from various features \cite{Kong2012_EEW_features, Kong2016_SciAdvances}. (2) The MyShake cloud server collects data from smartphones including state of health heartbeats, event triggers of the phones when they detect earthquake-like motions, and the corresponding time series of the phones' accelerometer data. A spatial and temporal clustering algorithm runs on the cloud server to aggregate information from multiple smartphones to identify earthquakes \cite{Kong2019_SRL_ML}. Whenever a phone detects an earthquake-like motion, it sends a trigger message including the time, location, and amplitude to the cloud server, where the clustering algorithm will confirm the earthquake and estimate earthquake parameters such as magnitude, location, and origin time. At the same time, the phone also records 5-minute (1 minute before and 4 minutes after the trigger) 3-component time series of acceleration and upload to the cloud server when the phone connects to WIFI and power. A detailed technical system architecture can be found in \cite{Kong2016_SciAdvances}.

\begin{table}
  \caption{Top 10 phone brands among 276,140 MyShake users.}
  \label{tab:freq}
  \begin{tabular}{c|c||c|c} 
    \toprule
    Phone Brand & Percentage & Phone Brand & Percentage\\ 
    \midrule
   Samsung & 43.5\% & Sony & 4.4\%\\
   Motorola & 6.1\% & Google & 4.2\%\\ 
   LG & 5.6\% & HTC & 3.5\%\\ 
   Verizon & 4.7\% & Xiaomi & 2.8\%\\ 
   Huawei & 4.5\% & Lenovo & 2.7\%\\ 
  \bottomrule
\end{tabular}
\label{tab:phone_types}
\end{table}

\begin{figure}
\includegraphics[width=0.48\textwidth]{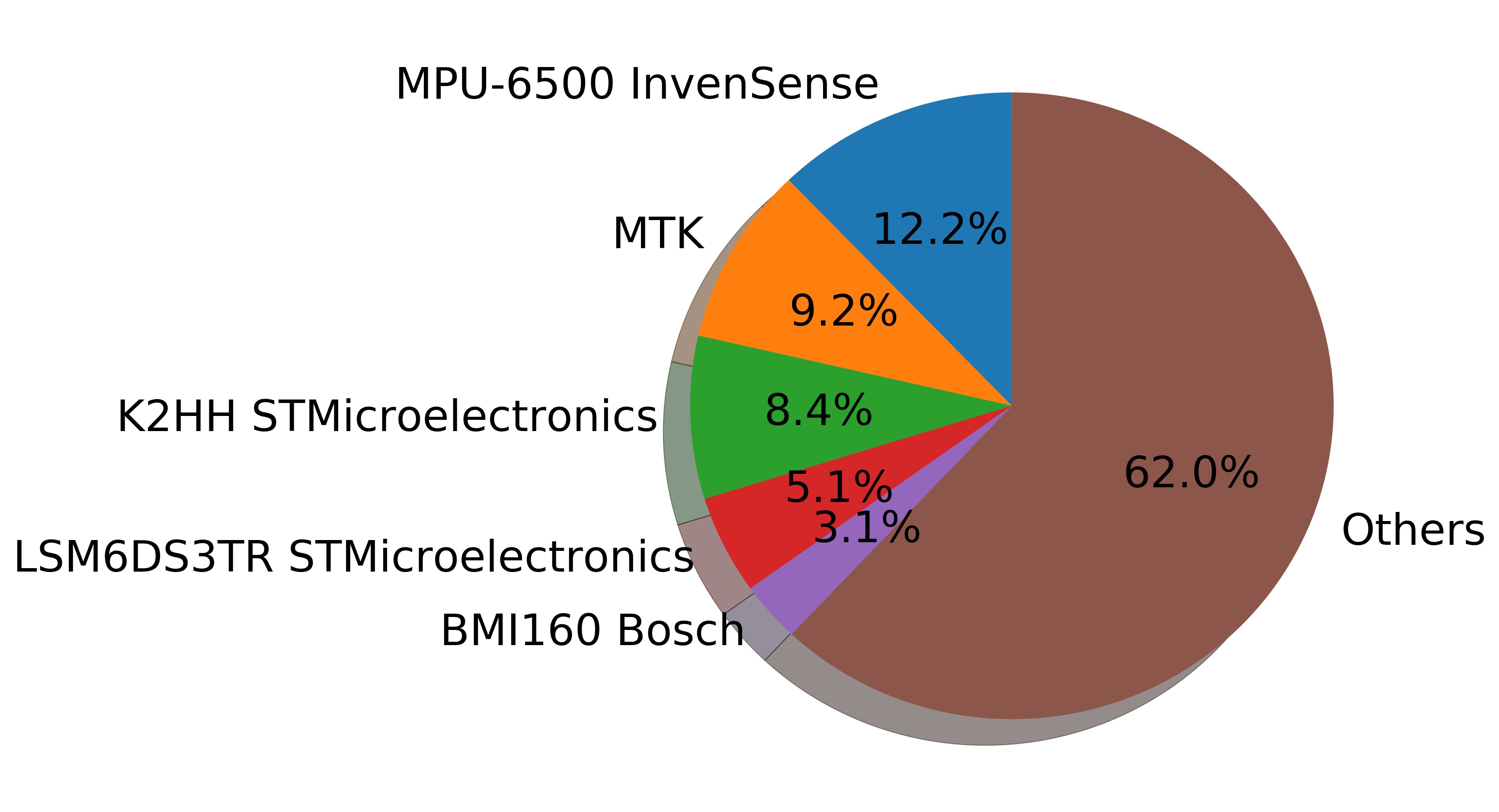}
\caption{Top 5 accelerometer types in MyShake users' phones. Data are from 276,140 users.}
\label{fig:acc_vendor}
\end{figure}

\section{EEW Challenges for MyShake}
While the current MyShake system is capable of detecting earthquakes on individual phones and collectively confirming earthquakes at the seismic network level, a number of unique challenges need to be addressed in order to issue real-world EEW using such a smartphone-based seismic network. The key is to concurrently achieve highly accurate earthquake detection and highly efficient early warning, which requires pushing the boundaries of prior research. In this section, we highlight two key challenges that we have been working on, and discuss a few other challenges that are relevant to the mobile computing community. 

\subsection{Diversity of Sensing Hardware}

Unlike traditional seismic networks, the MyShake network consists of individual users' smartphones. It is difficult to control the consistency of the sensing hardware. From the data collected by MyShake, it is clear that there exists a wide spectrum of brands/makes of the phones and sensors. As a result, these phones have different detection sensitivities due to the quality of the sensors, and even two different phones/sensors at the same location may or may not trigger on the same motion. A one-size-fits-all solution would not work well, and a comprehensive understanding of device diversity and sensitivity for earthquake detection is needed for the design of adaptive system strategies and parameters. The real-world data collected by MyShake include phone/sensor information as well as recorded waveforms from users' phones. An initial analysis using 276,140 MyShake users' data reveals a wide variety of phone brands and sensor types. The top 10 phone brands and their proportions are shown in Table~\ref{tab:phone_types}. The top 5 accelerometer types account for about 40\% of all the phones, as shown in Figure~\ref{fig:acc_vendor}, and there are in total 367 different types of sensors in MyShake users' phones.

\begin{figure}
\includegraphics[width=0.45\textwidth]{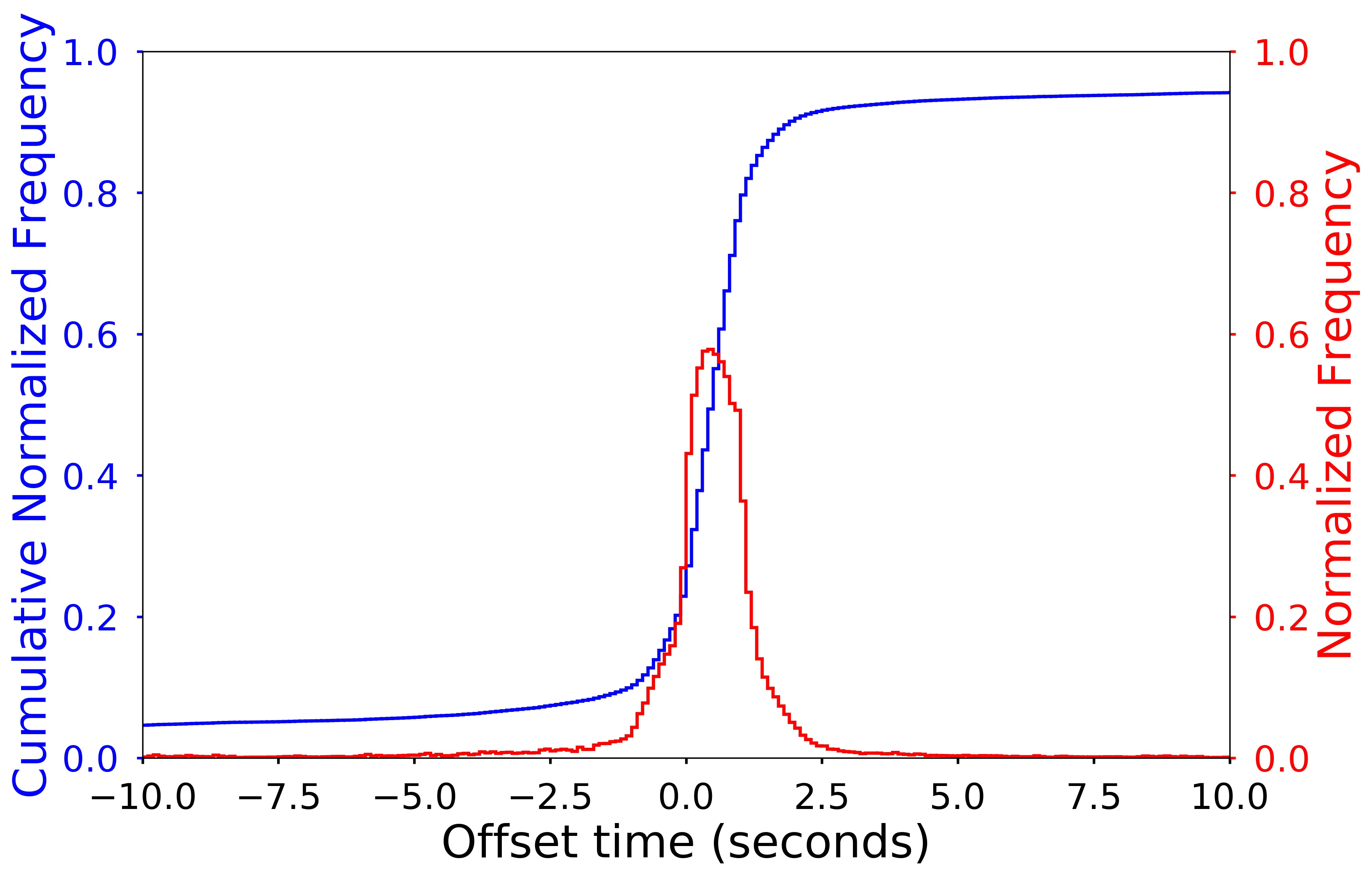}
\caption{Distribution of MyShake phones' time offset based on 1 million randomly sampled NTP records using hourly synchronization.}
\label{fig:offset_time}
\end{figure}

Inconsistent timing among commodity phones is particularly challenging for smartphone-based EEW. Seismic waves travel at about 3--6 km/s, and our clustering detection on the cloud server looks for coherent triggers from multiple phones. If there is a 5-sec offset on the phones with respect to the true time, seismic waves could have traveled for 30 km, which significantly impacts the effectiveness of earthquake detection and early warning. In the current MyShake system, each phone synchronizes via NTP (Network Time Protocol) every hour. Figure \ref{fig:offset_time} shows the distribution of time offset between the phone clock and the true time at the time of NTP synchronization for 1 million randomly sampled records from our database. We can see that using hourly synchronization, most of the phones would have an offset time within 2.5 seconds.

\subsection{Dynamics of Users and System}

\begin{figure}
\includegraphics[width=0.48\textwidth]{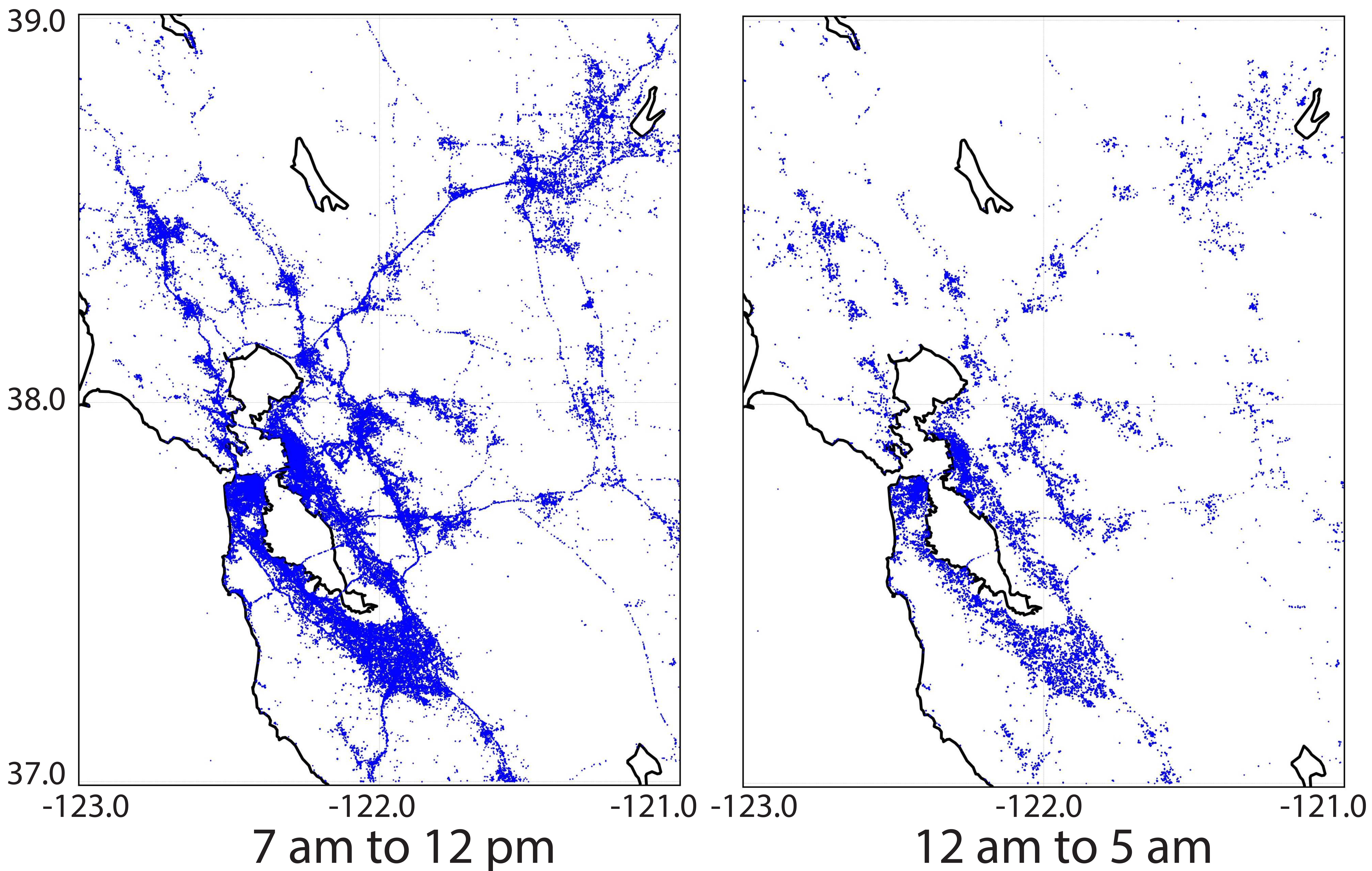}
\caption{MyShake user distribution (sampled every 2 hours) in the San Francisco Bay Area. (Left) During the day from 7am to 12pm; (Right) During the night from 12am to 5am. Data used here are from 2017-07-01 to 2018-07-01.}
\label{fig:network_configuration_spatial}
\end{figure}

\begin{figure}
\includegraphics[width=0.48\textwidth]{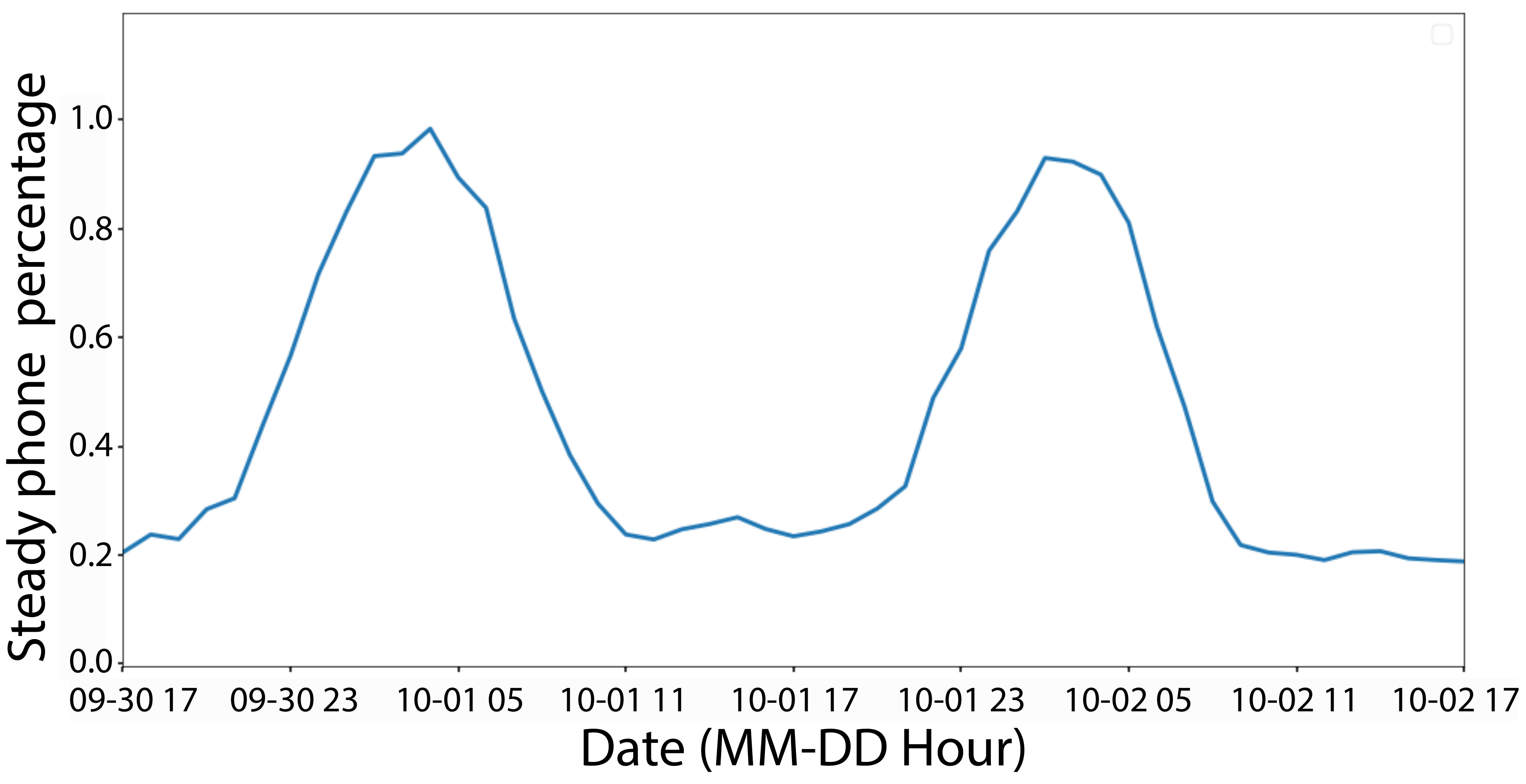}
\caption{Percentage of steady MyShake phones (steady for more than 30 minutes) in the San Francisco Bay Area. Data plotted here are from 2017-09-30 5pm to 2017-10-02 5pm.}
\vspace{-1.0em}
\label{fig:network_configuration_time}
\end{figure}

In the MyShake smartphone-based seismic network, the phones move with their users. As such, the seismic network changes constantly both in space and time, and the detection capability of the network varies by region and over time. For example, it is observed that more phones move to office (home) during the day (night), and during the night, more phones are stationary for longer periods of time. For example, Figure~\ref{fig:network_configuration_spatial} shows the spatial distribution of MyShake users in the San Francisco Bay Area during day and night. We can see clear network  distinctions between the two time periods, where the network is much denser and more spread out during the day vs. during the night. Figure~\ref{fig:network_configuration_time} shows the percentage of phones which were steady for more than 30 minutes in the same area for two consecutive days. Again, we see a wide fluctuation of the percentage over time.  When phones are steady, they are in the best positions to detect earthquakes, while phones on the move cannot detect earthquakes reliably. Therefore, the percentage of steady phones in an area at any given time is actually a good indicator of the network's detection capability. Depending on the specific time and region when an earthquake occurs, the system may have different numbers of phones available at different locations, and different earthquake detection strategies and parameters may be needed for the system to detect the specific earthquake. 

\subsection{Other Challenges}

Besides the two challenges mentioned above, there exist other challenges that are related to supporting EEW in the MyShake system. For instance, when an earthquake strikes, near real-time communication is crucial in terms of receiving trigger messages from individual phones and sending EEW messages to millions of users in one region (e.g., around 7 million people in the San Francisco Bay Area). Understanding the scalability and limitations of the current system and developing innovative techniques to reduce the notification latency could have a significant impact on saving lives and critical infrastructures. In addition, tradeoffs between the confidence and latency of earthquake detection should be carefully examined, and some multi-tiered EEW system design may be necessary.  Furthermore, given the disruptive nature of EEW, system security is paramount to avoid incidental or targeted attacks. One particular aspect of system security is related to spoofed earthquake triggers, such as how easy it is to spoof earthquake triggers, and how robust the system is against spoofed earthquake triggers from individual phones, ad-hoc or coordinated groups of phones in a specific region and time period. 

\section{Tackling the EEW Challenges}

We have conducted some preliminary research in order to tackle the challenges of diverse sensing hardware and user/system dynamics. Specifically, we analyze the large-scale real-world data collected via MyShake, and have designed and developed a simulation platform to model different phone/user/system properties and evaluate their impact on the performance of EEW. 

\begin{figure}
\includegraphics[width=0.45\textwidth]{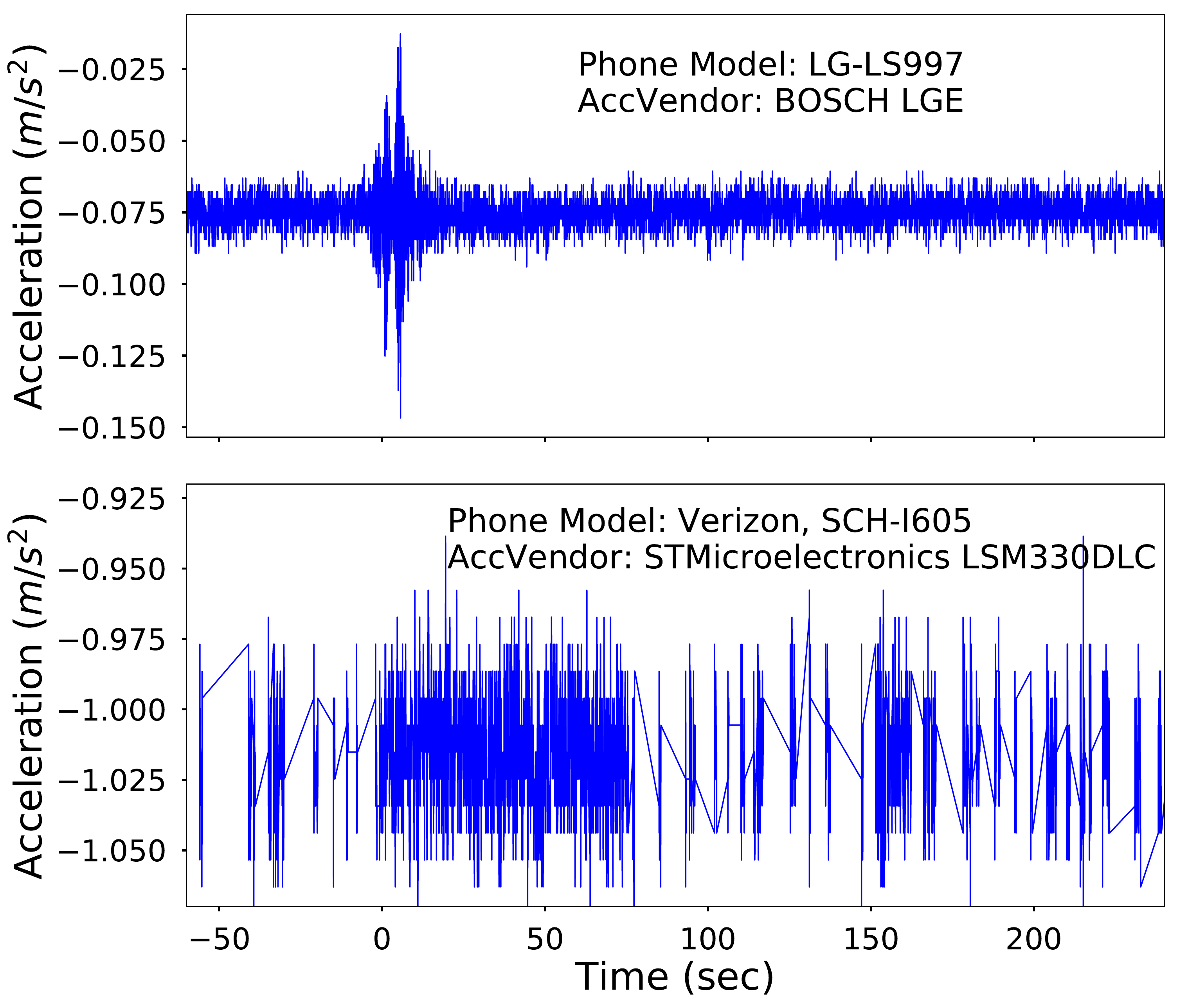}
\caption{Waveform examples from 2 different users. The top waveform has good quality and records an earthquake, while the bottom waveform has lots of missing data.}
\vspace{-1.0em}
\label{fig:waveform_example}
\end{figure}

\subsection{Tackling Diversity of Sensing Hardware}
Given the heterogeneity of the sensing hardware in the MyShake network, it is important to link sensor/phone types to the quality of motion waveforms. As mentioned earlier, the brand/model of the phones and accelerometers are collected by the MyShake app. Meanwhile, time series of the acceleration recorded by these phones are also collected. Data quality information can be extracted from the collected waveforms. For example, using the 1-minute waveform before each earthquake trigger (background noise), the noise level of the phone can be estimated by calculating the standard deviation of the noise. The waveforms also contain data gaps which usually appear for some users, and the relative occurrence of these glitches could be monitored and used as an indicator of the sensor quality for different users.  Besides, the sampling interval distribution can also tell us the recording stability of the sensor. Currently, we collect the 25th, 50th, 75th percentiles of the sampling intervals. Figure~\ref{fig:waveform_example} shows two example waveforms from two different users. We can see that the first waveform has good quality and adequately captures a real-world earthquake event, while the second waveform has lots of missing data. For the first waveform, the standard deviation of the noise level is around 0.005 $m/s^2$, with no data gaps that are larger than 1s, and the 25th, 50th, 75th percentiles of the sampling interval are 39, 40, 41 msec. In contrast, the bottom waveform has a standard deviation of 0.03 $m/s^2$, 35 instances of data gaps that are larger than 1s (32 of them are larger than 2s), and 1, 59, 60 for the 25th, 50th, 75th percentiles. Based on these extracted metrics from the waveforms, we can model the sensing quality for different phones/sensors/users. The sensing quality can then be used as a weighting function of sensor importance/confidence in the detection algorithm to downgrade sensors with poor quality. Since the detection algorithm is based on collective intelligence of many smartphones, this quality measure can help make the algorithm more adaptive. For instance, in a given region, if only low-quality sensors are triggered, more triggered phones may be needed to declare an earthquake so as to decrease the chance of issuing a false warning, and vice versa.  

With the current implementation of NTP synchronization every hour, most of the MyShake phones have a time accuracy within 2 seconds. We are investigating ways to further improve this accuracy. One approach is to increase the frequency of NTP synchronization, and the question is how smaller time intervals would improve time accuracy at the cost of increased overhead. Another approach is to utilize the time of arrival when the server receives heartbeats and triggers from individual phones, and the question is how location and neighboring phones may help augment existing time synchronization/calibration strategies.

\subsection{Tackling Dynamics of Users and Systems}
To tackle the challenge of user and system dynamics, we have developed a simulation platform to generate triggers caused by earthquakes in order to mimic actions in the MyShake seismic network and evaluate the performance of different design strategies. Our simulation platform works as follows. Given the information of a specific earthquake,  we first use global population density within 1 $km^2$ to sample MyShake users in the region.  Depending on the time of occurrence of the earthquake, different number of steady phones will be simulated. This sampling is based on the statistical relationship extracted from MyShake observations, which is shown in Figure 7b of \cite{Kong2019_SRL_ML}. Then, using both physical modeling (the spread of the P and S waves is based on a homogeneous medium, the Peak Ground Acceleration is based on an attenuation model developed by \cite{Cua2005}) and statistical modeling (learned from MyShake observations, the distribution of the time errors on the phones are shown in Figure~\ref{fig:offset_time}), we can determine each phone's triggering probability when different seismic waves pass by and the time of trigger for the specific earthquake with corresponding uncertainties. In addition, based on false positive triggers observed overtime (see Figure 2 in \cite{Kong2019_SRL_ML}), randomized false positive triggers and uncertainties in trigger time are added to the simulation. Finally, the triggers generated by the simulation platform can be evaluated by spatial-temporal clustering to confirm the earthquake and estimate its corresponding parameters. The current algorithm under testing is DBSCAN (Density-Based Spatial Clustering of Applications with Noise) \cite{Ester1996}, which is modified to accommodate the temporal information. 

\begin{figure}
\includegraphics[width=0.5\textwidth]{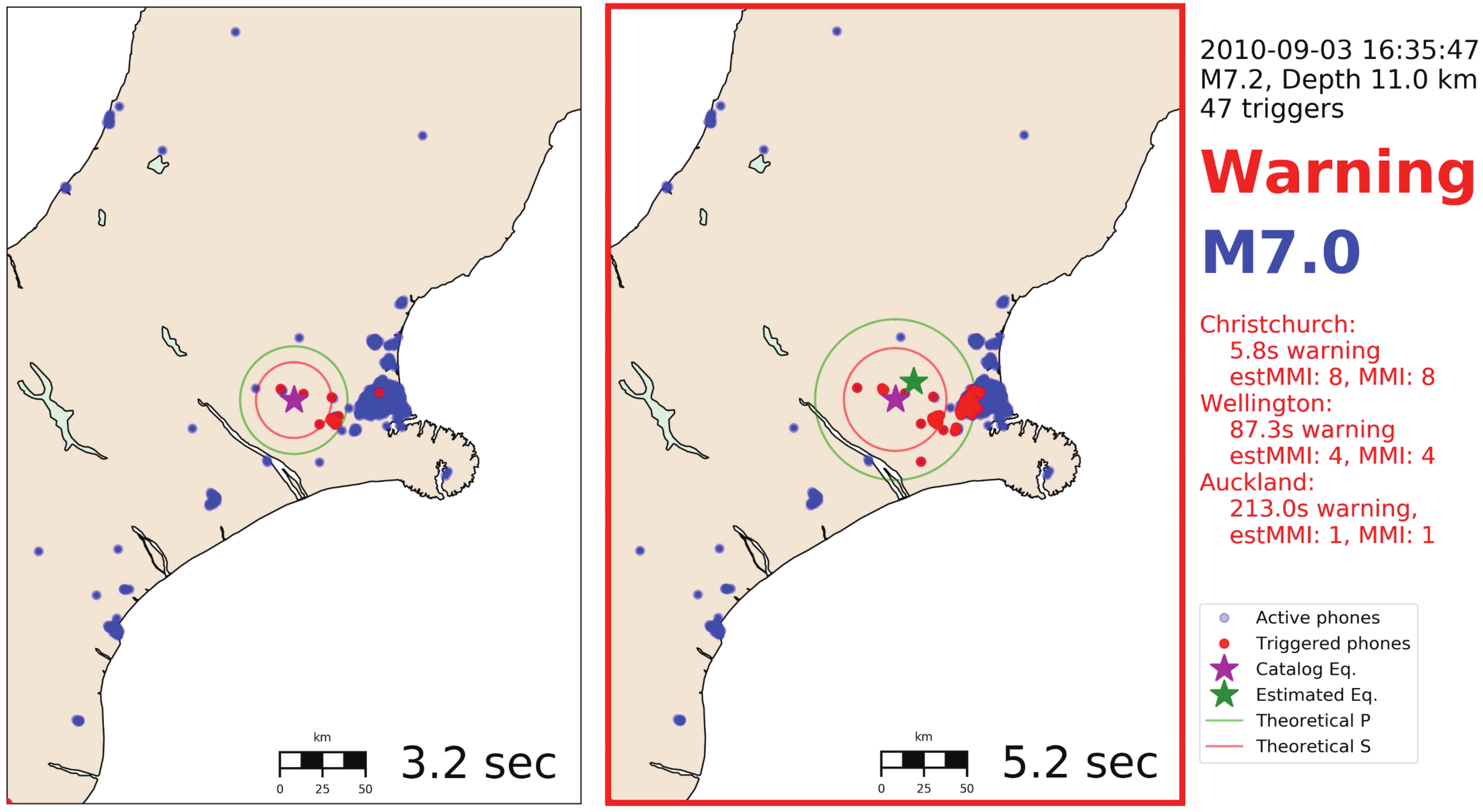}
\caption{Simulation snapshots for the 2010 New Zealand M7.2 Darfield earthquake, obtained at 3.2 seconds (left) and 5.2 seconds after the earthquake, respectively. The origin of the earthquake is indicated by the purple star. The legend on the right shows the time when the MyShake system detected the event, and estimated the magnitude as M7.0 (blue fonts) at the location indicated by the green star. The blue dots are steady phones running MyShake at the moment of the earthquake, and the red dots are phones triggered by the earthquake. The green and red circles show the two types of seismic waves -- P and S waves. The estimated intensity MMI (Modified Mercalli Intensity) and the warning time for 3 nearby cities are shown by the red text.}
\vspace{-1.0em}
\label{fig:NZ_simulation}
\end{figure}

Using this simulation platform, we can easily generate multiple simulations for different network configurations to reflect spatial and temporal changes of MyShake users according to the occurrence time of the earthquake, and evaluate the effectiveness of different strategies and parameters for adaptive earthquake detection and early warning. Figure \ref{fig:NZ_simulation} shows the result of one example simulation. We can see the propagation of the seismic waves and the corresponding phones that are triggered in the network. The triggers outside the green seismic wave circle are random false positive triggers. In this simulation, our system is able to detect the earthquake  within 5.2 seconds after the initial onset of the earthquake, and the green star in the figure shows the estimated location.

\section{Related Work}
Besides MyShake, there are multiple efforts to develop EEW systems using smartphone sensors in the seismology community. \cite{Faulkner2011} talked about a smartphone EEW system, but there was no public deployment. \cite{Finazzi2016} also aims to build a global smartphone early warning system, but it lacks the capability of MyShake to separate earthquake signals from other human activities. There is also an app that detects earthquakes by monitoring when users launch the app and collecting users' reports \cite{Bossu2018}, but it is much slower in terms of detection due to the added human reaction time. Our own prior MyShake publications were in the seismology domain and focused mostly on the functionalities and applications in geophysics. Specifically, \cite{Kong2016_SciAdvances, Kong2012_EEW_features} described the initial development of the system and the design of the core ANN algorithm. \cite{Kong2016_GRL} and \cite{Kong2018_SRL_Building} reported some seismological observations and the potential use of MyShake to monitor the health of  buildings. \cite{Kong2019_SRL_ML} described the machine learning algorithms used in the MyShake system. In contrast, this paper focuses on new systems challenges related to issuing real-world EEW and aims to seek expert feedback from the mobile computing community.  

Beyond the geophysics community, many efforts related to understanding sensing hardware heterogeneity including mobile devices and other low-cost sensors can provide insights to the MyShake project. \cite{Evans2014-js} investigated  performance of several low-cost accelerometers in terms of recording motions in the laboratory environment.  \cite{Stisen2015-lj} evaluated sensor biases, sampling rate heterogeneity and instability using 36 different devices, as well as their impact on the performance of human activity recognition. These studies used only a few models in controlled environments, and there was no corresponding evaluation in large-scale real-world applications. Researchers have also investigated sensor calibration in mobile devices, such as a time-varying Kalman filtering calibration technique to reduce sensor biases~\cite{Batista2011-rd} and a machine learning based multiposition calibration scheme to address hardware heterogeneity in mobile devices~\cite{Grammenos2018-bf}. In our work, using the large-scale real-world data collected via MyShake, we will evaluate sensing quality in terms of measuring ground motion, and further leverage/develop sensor calibration techniques to improve sensing quality and earthquake detection accuracy. 

Characterizing human mobility dynamics using various datasets has received considerable attention. \cite{Kim2006-oy} extracted a human mobility model using 13-month wireless network traces collected from WiFi APs at Dartmouth College. \cite{Hsu2007-rj} used WLAN traces to create a time-variant community mobility model.  \cite{Gonzalez2008-do} derived a universal model to explain how individuals move using cellular network data in a European country. \cite{Isaacman2012-ji} proposed an approach to model how large populations move within different metropolitan areas using Call Detail Records. All these works aim to model human movement as a spatial-temporal relationship. Our work builds upon human mobility analysis, but further considers spatial-temporal availability and dynamics of steady phones for effective and efficient earthquake detection at both the individual phone level and overall seismic network level. 

\section{Conclusions and Future Directions}

Earthquakes are serious hazards globally, and MyShake has demonstrated the feasibility of building a smartphone-based seismic network for earthquake detection and early warning at the global scale. The initial deployment of MyShake has been successful, generating valuable data and new insights. In this paper, we have highlighted two key challenges for real-world EEW, namely, sensing heterogeneity and user/system dynamics, and potential solutions that we are exploring in terms of sensing quality measure and a simulation platform to model phone/user/system and adapt to different earthquake scenarios. Further improvements of our work include adaptive algorithms that take into account sensing quality and user/system dynamics, as well as simulations and real-world evaluations of those algorithms. 

This paper is our first step towards connecting with the mobile computing community. Several EEW challenges remain for real-world deployment, such as EEW system scalability, latency and security, which can really benefit from the expertise of the mobile computing community. Furthermore, while our current focus is issuing earthquake early warning to the public, we envision much broader use of MyShake and smartphone-based seismic network from the hazard preparation and response aspects in smart cities. Specifically, a system like MyShake could be used before, during and after earthquake events, such as proactive structural surveillance, risk assessment, and context-aware earthquake education before earthquakes occur, EEW during an earthquake, as well as emergency response,  rapid hazard information distribution, and long-term learning after an earthquake.

\begin{acks}
We would like to thank the shepherd Xia Zhou and reviewers of our paper for their insightful and constructive feedback. 
  MyShake is a joint collaboration between the Berkeley Seismology Laboratory and Deutsche Telecom Silicone Valley Innovation Center. The Gordon and Betty Moore Foundation funded this project through grant GBMF5230. This work was also supported in part by the National Science Foundation under grant number 1442971. Finally, we thank the MyShake team members and all the MyShake users! 

\end{acks}

\bibliographystyle{ACM-Reference-Format}
\bibliography{sample-bibliography}

\end{document}